\documentclass[aps,prl,twocolumn,groupedaddress,showpacs]{revtex4}
\usepackage{graphicx}
\usepackage{dcolumn}

\begin{document}


\title{Liquid Crystal Foams: Formation and Coarsening}


\author{Mark Buchanan}
\email[]{buchanan@nat.vu.nl}
\altaffiliation{Vrije Universiteit, Amsterdam, The Netherlands.} 
\affiliation{Centre de Recherche Paul-Pascal, CNRS, \\ 
Avenue du Docteur-Schweitzer, F-33600 Pessac, France.}


\date{\today}

\begin{abstract}
Coarsening in foams made from the pure liquid crystal, 8CB, has been
studied. The foam was made in the nematic phase ($\textrm{T} = 35
^{\circ}\textrm{C}$) by bubbling nitrogen through the pure liquid crystal. The
coarsening behavior was investigated at three temperatures; at $\textrm{T} = 22
^{\circ}\textrm{C}$ and $33^{\circ}\textrm{C}$ in the smectic phase and at
$\textrm{T} = 34^{\circ}\textrm{C}$ in the nematic phase. In smectic and
nematic phases the mean bubble radius $\langle R \rangle$ has been measured as
a function of time $\langle R \rangle \sim t^{\lambda}$. In classical wet soap
foams the growth exponent is typically $\lambda \approx 0.33$ where coarsening
is by gas diffusion from bubbles with high curvature to bubbles with low
curvature. In liquid crystal foams a growth exponent, $\lambda = 0.20 \pm 0.05$
is observed. This may be explained by the presence of defects at the surface of
the bubbles which slow down the coarsening behaviour. This growth exponent can
be observed in both nematic and smectic phases. At higher temperatures
typically $>35^{\circ}\textrm{C}$ coalescence dominates the coarsening
behaviour. In the isotropic state, $>41.5^{\circ}\textrm{C}$, the foam is
rapidly unstable.
\end{abstract}


\maketitle

The dynamic behaviour of classical foams which are usually made with an aqueous
solution of surfactant have been well studied. Coarsening, drainage,
structure and rheology are of fundamental interest in the study of
foams~\cite{stavans:cellular,durian:foamsrev,rieser:gasdiff,magrabi:coarsening,bisperink:beer,monnereau:coarsening}.
In this article we are primarily interested in the coarsening behaviour where
the temporal evolution of the mean bubble size is expressed as $\langle R
\rangle \sim t^{\lambda}$ where $t$ is the time and $\lambda$ is the growth
exponent. When coarsening is dominated by gas diffusion between bubbles the
growth exponent depends on the liquid fraction of the
foam\cite{hilgenfeldt:coarsening,cheng:diffusion,lemlich:diffusion,devries:bubbles,stavans:cellular}.

In dry foams, bubbles are polyhedral-like and separated by flat films and the
growth exponent is $\lambda \approx \frac{1}{2}$ due to gas diffusion through
the films\cite{liftshitz:ripe2,wagner:ripe}.  In contrast in wet foams bubbles
appear to be spherical and coarsening occurs by gas diffusion through the three
dimensional continuum of the liquid phase. The growth exponent is slower and an
exponent $\lambda \approx \frac{1}{3}$ is
observed\cite{kabalnov:ostrip,egelhaaf:ostrip}. Coarsening can also occur by
film rupture leading to rapid coalescence of the bubbles and a growth exponent
of $\lambda \approx 1$ \cite{burnett:breaking}.

In this paper we present a study of the coarsening behaviour of a foam made
from a pure liquid crystal where neither solvent nor surfactant is
present. Despite much previous work on soap foams there has never been any
studies performed on foams which are made from a pure thermotropic liquid
crystal.

An important difference between a classical foam and a liquid crystal foam is
the role of elastic distortions due to defects near the
surface\cite{oswald:bubble,zywocinski:edge}.  Recent studies on films made from
liquid crystals show that edge dislocations in a smectic film meniscus can
influence its shape~\cite{picano:meniscus,picano:disjoining}. A coupling
between defects and the surface energy of the film can influence the type of
defect observed. This interaction exists due to the elastic distortion field
exerted by the defect on the film surface.

In this study the foam was made from pure 8CB (4-n-octylcyanobiphenyl) liquid
crystal. Below $\textrm{T} = 21.5 ^{\circ}\textrm{C}$ the liquid crystal is in
the solid phase. The smectic phase exists between $\textrm{T} = 21.5
^{\circ}\textrm{C}$ and $33.4^{\circ}\textrm{C}$ and the nematic phase
exists between $\textrm{T} = 33.4 ^{\circ}\textrm{C}$ and
$41.5^{\circ}\textrm{C}$. Above $\textrm{T} = 41.5^{\circ}\textrm{C}$ the
liquid crystal is an isotropic phase.

\begin{figure}[!t]
\includegraphics[width=8cm]{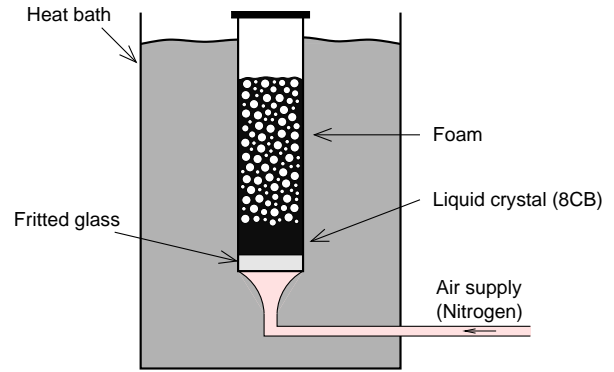}
\caption{\label{fig1} Experimental Setup: The foam is contained in glass cell
which is immersed in a temperature controlled heat bath.}
\end{figure}

\begin{figure}[!t]
\includegraphics[width=8cm]{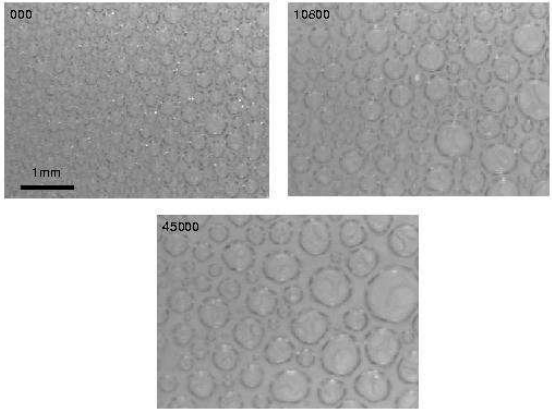}
\caption{\label{fig2} Images of bubble growth in a smectic phase $\textrm{T} = 33 ^{\circ}\textrm{C}$. Times are
displayed in seconds.}
\end{figure}

All foams were made in a glass cell with dimensions 1cm $\times$ 1cm $\times$
10cm. The cell was immersed in a temperature controlled heat bath
(Fig.~\ref{fig1}). To make the foam the liquid crystal must first be heated
into the nematic phase ($\textrm{T} = 35 ^{\circ}\textrm{C}$).  Then, nitrogen
gas is passed through fritted glass at the base of the cell to form the foam
until a sufficient amount is made ($\sim 6$cm high). After making the foam the
temperature is adjusted appropriately for the experiment. Then the evolution of
bubble growth at the glass surface is followed with a time-lapse video recorder
and a frame-grabbing PC (Fig.~\ref{fig2}). 

The evolution of the bubble growth was followed by observing the foam at the
glass surface.  Since the surface only represents a 2D slice of the foam
small bubbles are under-counted. The bubble distribution can be corrected in
order to obtain the true 3D bubble size distribution~\cite{cheng:errors}. The
mean bubble size can be determined using $\langle R \rangle = N (\sum_i
r_i^{-1} )^{-1}$ where N is the total number of bubbles measured each with
radius $r_i$ \cite{a:bubble}. This method to calculate the mean size was
used for all experiments in this study. 

The liquid volume fraction of the foam after fabrication can be roughly
determined ($\pm 10\%$) by rapidly heating the foam and measuring the liquid
recovered from a foam of known volume.  The measured liquid fraction
immediately after fabrication is approximately $\phi_l = 0.40$. It is easily
observed from images that the bubbles are initially round and there is $\sim
50\mu \textrm{m}$ of distance between bubbles.  The liquid fraction was 
measured at the end of the experiments and was observed to reduce to about
$\phi_l \sim 0.34$.  Throughout the coarsening the film thickness, i.e. the
distance between bubbles, is approximately constant.  In the nematic phase the
foam is drier and hence bubbles are closer to each other than in the smectic
phase.

The temporal evolution of the mean bubble radius was measured in smectic and
nematic foams. In the smectic phase growth exponents were determined at
$\textrm{T} = 22 ^{\circ}\textrm{C}$ and $\textrm{T} = 33 ^{\circ}\textrm{C}$.
In the nematic phase the growth exponent was measured at $\textrm{T} = 34
^{\circ}\textrm{C}$. In order to determine the growth exponent a log-log plot
is generated. The mean radius varied over time such that $\langle R(t) \rangle
\sim t^\lambda$. An exponent was determined from each run and an average
exponent, $\langle \lambda \rangle$ was calculated. 

\begin{figure}[!h]
\begin{center}
	\includegraphics[width=8cm]{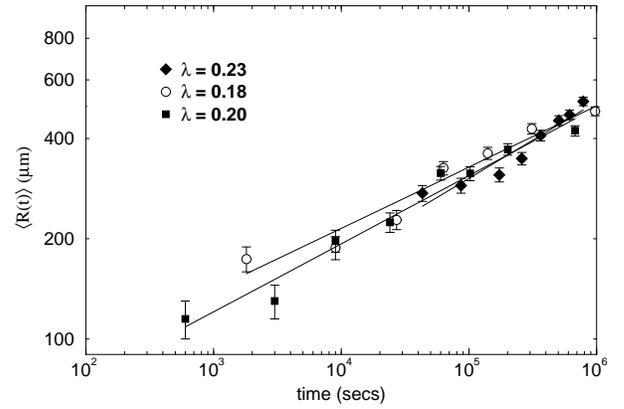} \\
\caption{Mean bubble size as a function of time in smectic phase
$\textrm{T} = 22 ^{\circ}\textrm{C}$. Each symbol represents a separate
experiment and the line is a fit to all data points.} 
\label{fig3}
\end{center}
\end{figure}

\begin{figure}[!h]
\begin{center}
	 \includegraphics[width=8cm]{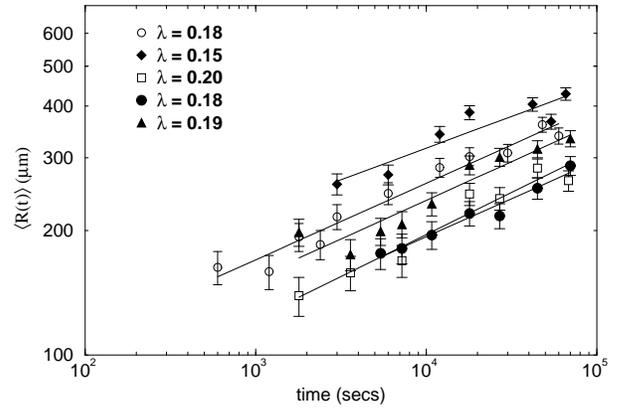}\\
\caption{Mean bubble size as a function of time in smectic phase $\textrm{T} =
33 ^{\circ}\textrm{C}$.} 
\label{fig4}
\end{center}
\end{figure}

\begin{figure}[!h]
\begin{center}
	\includegraphics[width=8cm]{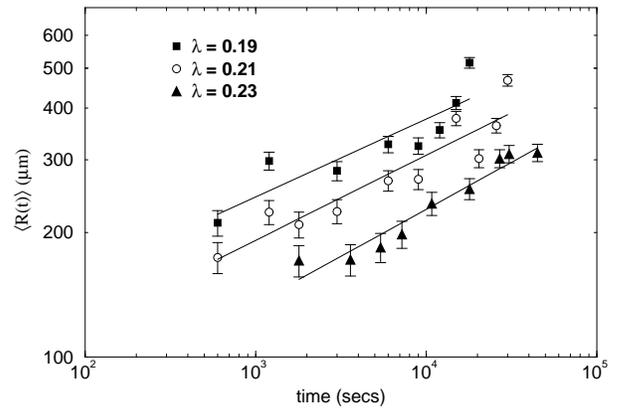}\\
\caption{Mean bubble size as a function of time in nematic phase $\textrm{T} =
34 ^{\circ}\textrm{C}$.}
\label{fig5}
\end{center}
\end{figure}

Figures~\ref{fig3} - \ref{fig5} show the temporal evolution of the mean
bubble size for different runs, represented by different symbols on the graph,
for $\textrm{T} = 22^{\circ}\textrm{C}$, $33^{\circ}\textrm{C}$ and
$34^{\circ}\textrm{C}$. 
Large variations between runs is attributed to the
difference in initial size distribution of the bubbles during preparation.

In the smectic phase at low temperature ($\textrm{T} = 22^{\circ}\textrm{C}$)
the bubble growth was followed for 10 days. The best fits of the data for each
run are plotted and the mean growth exponent averaged over all the runs are
$\langle \lambda \rangle \approx 0.20\pm0.05$.  In the smectic phase at high
temperature ($\textrm{T} = 33^{\circ}\textrm{C}$) the mean bubble size is
measured over 1 day. Again, several runs are represented by different symbols
and a best fit plotted for each run. These results give an average mean
growth exponent,$\langle \lambda \rangle \approx 0.18 \pm 0.05$.  In the
nematic phase ($\textrm{T} = 34^{\circ}\textrm{C}$) the growth exponent is
measured to be $\langle \lambda \rangle = 0.21 \pm 0.05$.  At all three
temperatures we observe a growth exponent close to one fifth.

At temperatures between $36^{\circ}\textrm{C}$\
\raisebox{-0.2ex}{$\stackrel{\scriptstyle<}{\scriptstyle\sim}$}\ T\
\raisebox{-0.2ex}{$\stackrel{\scriptstyle<}{\scriptstyle\sim}$}\  $41.5^{\circ}\textrm{C}$
the foam is very unstable and the main process responsible for the bubble
growth is by rapid coalescence. If the foam is heated above
$41.5^{\circ}\textrm{C}$ it instantly collapses. This behaviour can be
exploited when we require to destroy all of the foam before we make a new foam
for the next experiment.
 
\begin{figure}
\includegraphics[width=8cm]{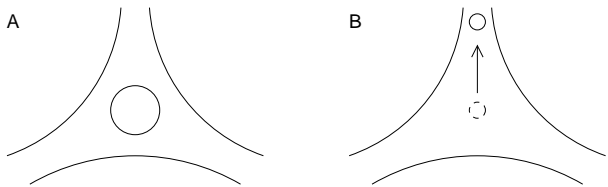}
\caption{\label{fig6}(A) Bubble trapped in the node. (B) Bubble shrinks and
escapes upward through the films.}
\end{figure}

The dynamics of small bubbles has also been followed in smectic and nematic
foams. In a smectic foam the small bubbles shrink and eventually disintegrate
while remaining confined in the foam. However in the nematic phase small
bubbles shrink and escape upon reaching a size of about $20-40\mu \textrm{m}$
(figure~\ref{fig6}).  They then rise to the top of the foam through the
Plateau borders.

\begin{figure}[!h]
\includegraphics[width=8cm]{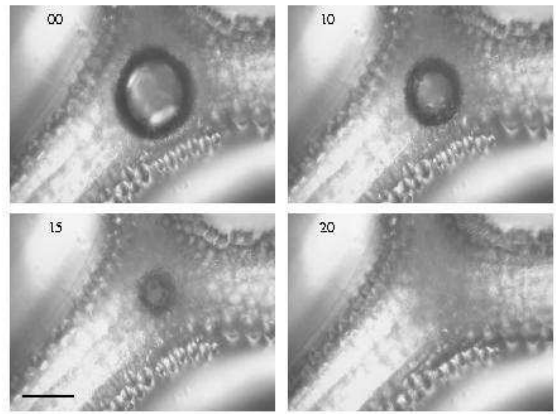}
\caption{\label{fig7} Small bubble trapped between three large bubbles
shrinks over time. Scale bar is $50\mu\textrm{m}$ and time shown in seconds.}
\end{figure}

By removing the foam from the cell and placing it between a glass slide and
a coverslip we were able to observe a 2D foam. The cell thickness was 
approximately $50\mu$m which allowed visualization of the bubble surface. In
this geometry, defects at the surface of the bubbles were observed in a
microscope with partially crossed polars. The time-evolutution of a small
bubble shrinking in a 2D foam is shown in figure~\ref{fig7}.

Above we have shown that liquid crystal foams can be prepared, and that they do
not have the typical $t^{1/3}$ growth behavior observed in classical wet
foams. For liquid crystal foams, growth exponents $\lambda \sim 0.20$ have been
measured.

In the classical theoretical treatment, coarsening in foams follows from
Lifshitz, Slezov and Wagner theory
\cite{liftshitz:ripe2,wagner:ripe}. Diffusion of gas is driven from small
bubbles with higher excess chemical potential ($\Delta \mu(r) = \mu_{\infty}
-\mu(r)$) to larger bubbles.  The higher chemical potential is due to high
curvature of smaller bubbles which leads to a higher concentration $c(r)$ of
gas being dissolved at the bubble surface
\[
c(r) = c_{\infty} \exp \left(\frac{\Delta \mu(r)}{RT}\right), 
\hspace*{5mm}
\textrm{where}
\hspace*{5mm}  
\Delta \mu(r) = 2 V_m \frac{\sigma}{r}
\]
where $c_{\infty}$ is the bulk solubility of the gas, $V_m$ is the molar
volume, $R$ is the gas constant, $T$ is the temperature and $\sigma$ is the
surface tension.  Combining these two equation and linearizing we obtain
\[
c(r) = c_{\infty}\left(1 + \frac{2\sigma}{r} \frac{V_m}{RT} \right)
\]
Using Ficks law with linear approximation for the concentration gradient near
to the bubble surface we obtain an expression for the gas flux $j=D (\bar{c}
- c(r))/r$ which can be expressed in terms of the
bubble size and the mean radius, $r_c$ where $\bar{c} = c(r_c)$, 
\[
\frac{\textrm{d}r}{\textrm{d}t} = \frac{\sigma C}{r} \left[\frac{1}{r_c} -
\frac{1}{r} \right],
\hspace*{5mm} \textrm{where}
\hspace*{5mm}  
C = \frac{8Dc_{\infty}V_m}{9\rho RT} 
\]
and $D$ is the diffusion coefficient of the gas in the solvent, $\rho$ is the
density. From this the temporal evolution of the mean bubble radius in the foam
can be determined
\[
\frac{\textrm{d}r_{c}^3}{\textrm{d}t} = C\sigma
\]
For the case of classical soap foam the surface tension is independent of the
bubble size. If the energy per unit area on a bubble surface were to depend on
$R$ such that $\sigma = A/R^2$, where $A$ is a constant, then a $t^{1/5}$
growth law could be predicted. A possible source of an $R$ dependent surface
tension may come from the surface/defect interactions. For a liquid crystal
foam the surface tension can be modified by the influence of defects near to
the surface.

As bubbles shrink the elastic interaction of the defects with the surface
would intensify due to an increase in defect density (number of defects per
unit area) and hence leads to an increase in surface tension. This would be
reasonable if the number of defects at the surface of the bubble remained
constant. The defects at the bubble surface could either be giant edge
dislocations or focal-conic defects. A change in the defect type is unlikely to
lead to any significant change in the distortion field which affects the
surface energy of the bubble. From these surface/defect interactions
an R dependent surface tension can be assumed which may explain the observed
$t^{1/5}$ growth law. This is the first time such a scaling law has been
observed in the coarsening of foams. Furthermore this growth is significantly
slower than other calculated growth exponents.

Although the coarsening may be explained from liquid crystal properties there
are open questions regarding the stability and foamability. The stability of a
foam will generally depend on its ability to avoid rapid
coalescence, normally at temperatures close to the foaming temperature
$\textrm{T} = 36^{\circ}\textrm{C}$. When above this temperature bubbles
rapidly coalesce and the foam collapses. One possible explanation could be that
the viscosity reduces significantly to allow rapid drainage to dry the foam and
hence coalescence is more possible. Another possible explanation is that the
number of smectic layers at the bubble surface may also play a role in the foam
stability. As the temperature is increased the thickness of the smectic skin
reduces~\cite{picano:disjoining}.  If the width of the smectic skin is
sufficiently reduced then the gravitational effects could become significant in
inducing coalescence of the bubbles.

In the field of liquid crystals, free standing films and single film
bubbles~\cite{oswald:bubble} made from pure liquid crystal have been studied.
The ability to make a pure liquid crystal foam opens many doors for studying
other foam properties such as drainage, structure and rheology in this novel
system.

\begin{acknowledgments}
I would like to acknowledge P. Poulin, P. Barois and D. Roux for their kind
hospitality and support throughout my time at the CRPP and M. Cates, J. Leng,
F. Nallet, J. Bibette, P. Oswald, I. Stewart, N. Mottram and D. Head for useful
discussions. 
\end{acknowledgments}


\begin{thebibliography}{21}
\expandafter\ifx\csname natexlab\endcsname\relax\def\natexlab#1{#1}\fi
\expandafter\ifx\csname bibnamefont\endcsname\relax
  \def\bibnamefont#1{#1}\fi
\expandafter\ifx\csname bibfnamefont\endcsname\relax
  \def\bibfnamefont#1{#1}\fi
\expandafter\ifx\csname citenamefont\endcsname\relax
  \def\citenamefont#1{#1}\fi
\expandafter\ifx\csname url\endcsname\relax
  \def\url#1{\texttt{#1}}\fi
\expandafter\ifx\csname urlprefix\endcsname\relax\def\urlprefix{URL }\fi
\providecommand{\bibinfo}[2]{#2}
\providecommand{\eprint}[2][]{\url{#2}}

\bibitem[{\citenamefont{Stavans}(1993)}]{stavans:cellular}
\bibinfo{author}{\bibfnamefont{J.}~\bibnamefont{Stavans}},
  \bibinfo{journal}{Rep. Prog. Phys.} \textbf{\bibinfo{volume}{56}},
  \bibinfo{pages}{733 } (\bibinfo{year}{1993}).

\bibitem[{\citenamefont{Durian and Weitz}(1994)}]{durian:foamsrev}
\bibinfo{author}{\bibfnamefont{D.}~\bibnamefont{Durian}} \bibnamefont{and}
  \bibinfo{author}{\bibfnamefont{D.~A.} \bibnamefont{Weitz}}, in
  \emph{\bibinfo{booktitle}{Kirk-Othmer Encyclopedia of Chemical Technology}},
  edited by \bibinfo{editor}{\bibfnamefont{J.}~\bibnamefont{Kroschwitz}}
  (\bibinfo{publisher}{Wiley}, \bibinfo{address}{New York},
  \bibinfo{year}{1994}), vol.~\bibinfo{volume}{11}, pp. \bibinfo{pages}{783 --
  805}, \bibinfo{edition}{4th} ed.

\bibitem[{\citenamefont{Rieser and Lemlich}(1988)}]{rieser:gasdiff}
\bibinfo{author}{\bibfnamefont{L.~A.} \bibnamefont{Rieser}} \bibnamefont{and}
  \bibinfo{author}{\bibfnamefont{R.}~\bibnamefont{Lemlich}},
  \bibinfo{journal}{Journal of Colloid and Interface Science}
  \textbf{\bibinfo{volume}{123}}, \bibinfo{pages}{299 } (\bibinfo{year}{1988}).

\bibitem[{\citenamefont{Magrabi et~al.}(1999)\citenamefont{Magrabi,
  Dlugogorski, and Jameson}}]{magrabi:coarsening}
\bibinfo{author}{\bibfnamefont{S.~A.} \bibnamefont{Magrabi}},
  \bibinfo{author}{\bibfnamefont{B.~Z.} \bibnamefont{Dlugogorski}},
  \bibnamefont{and} \bibinfo{author}{\bibfnamefont{G.~L.}
  \bibnamefont{Jameson}}, \bibinfo{journal}{Chem. Eng. Sci.}
  \textbf{\bibinfo{volume}{54}}, \bibinfo{pages}{4007 } (\bibinfo{year}{1999}).

\bibitem[{\citenamefont{Bisperink et~al.}(1992)\citenamefont{Bisperink,
  Rolteltap, and Prins}}]{bisperink:beer}
\bibinfo{author}{\bibfnamefont{C.~G.~J.} \bibnamefont{Bisperink}},
  \bibinfo{author}{\bibfnamefont{A.~D.} \bibnamefont{Rolteltap}},
  \bibnamefont{and} \bibinfo{author}{\bibfnamefont{A.}~\bibnamefont{Prins}},
  \bibinfo{journal}{Advances in colloid and interface science}
  \textbf{\bibinfo{volume}{38}}, \bibinfo{pages}{13 } (\bibinfo{year}{1992}).

\bibitem[{\citenamefont{Monnereau and
  Vignes-Adler}(1998)}]{monnereau:coarsening}
\bibinfo{author}{\bibfnamefont{C.}~\bibnamefont{Monnereau}} \bibnamefont{and}
  \bibinfo{author}{\bibfnamefont{M.}~\bibnamefont{Vignes-Adler}},
  \bibinfo{journal}{Phys. Rev. Lett.} \textbf{\bibinfo{volume}{80}},
  \bibinfo{pages}{5228} (\bibinfo{year}{1998}).

\bibitem[{\citenamefont{Hilgenfeldt et~al.}(2001)\citenamefont{Hilgenfeldt,
  Koehler, and Stone}}]{hilgenfeldt:coarsening}
\bibinfo{author}{\bibfnamefont{S.}~\bibnamefont{Hilgenfeldt}},
  \bibinfo{author}{\bibfnamefont{S.~A.} \bibnamefont{Koehler}},
  \bibnamefont{and} \bibinfo{author}{\bibfnamefont{H.~A.} \bibnamefont{Stone}},
  \bibinfo{journal}{Phys. Rev. Lett.} \textbf{\bibinfo{volume}{86}},
  \bibinfo{pages}{4704} (\bibinfo{year}{2001}).

\bibitem[{\citenamefont{Cheng and Lemlich}(1985)}]{cheng:diffusion}
\bibinfo{author}{\bibfnamefont{H.~C.} \bibnamefont{Cheng}} \bibnamefont{and}
  \bibinfo{author}{\bibfnamefont{R.}~\bibnamefont{Lemlich}},
  \bibinfo{journal}{Ind. Eng. Chem. Fundam.} \textbf{\bibinfo{volume}{24}},
  \bibinfo{pages}{44 } (\bibinfo{year}{1985}).

\bibitem[{\citenamefont{Lemlich}(1978)}]{lemlich:diffusion}
\bibinfo{author}{\bibfnamefont{R.}~\bibnamefont{Lemlich}},
  \bibinfo{journal}{Ind. Eng. Chem. Fundam.} \textbf{\bibinfo{volume}{17}},
  \bibinfo{pages}{89 } (\bibinfo{year}{1978}).

\bibitem[{\citenamefont{De~Vries}(1972)}]{devries:bubbles}
\bibinfo{author}{\bibfnamefont{A.~J.} \bibnamefont{De~Vries}}, in
  \emph{\bibinfo{booktitle}{Absorptive bubble separation techniques}}, edited
  by \bibinfo{editor}{\bibfnamefont{L.}~\bibnamefont{R.}}
  (\bibinfo{publisher}{Academic Press}, \bibinfo{address}{New York},
  \bibinfo{year}{1972}), chap.~\bibinfo{chapter}{2}, pp. \bibinfo{pages}{7 --
  31}.

\bibitem[{\citenamefont{Liftshitz and Slezov}(1959)}]{liftshitz:ripe2}
\bibinfo{author}{\bibfnamefont{I.}~\bibnamefont{Liftshitz}} \bibnamefont{and}
  \bibinfo{author}{\bibfnamefont{V.}~\bibnamefont{Slezov}},
  \bibinfo{journal}{Soviet Physics J.E.P.} \textbf{\bibinfo{volume}{62}},
  \bibinfo{pages}{331} (\bibinfo{year}{1959}).

\bibitem[{\citenamefont{Wagner}(1961)}]{wagner:ripe}
\bibinfo{author}{\bibfnamefont{C.}~\bibnamefont{Wagner}},
  \bibinfo{journal}{Ber. Bunsenges.} \textbf{\bibinfo{volume}{65}},
  \bibinfo{pages}{58} (\bibinfo{year}{1961}).

\bibitem[{\citenamefont{Kabalnov et~al.}(1990)\citenamefont{Kabalnov, Makarov,
  Pertsov, and Shckukin}}]{kabalnov:ostrip}
\bibinfo{author}{\bibfnamefont{A.}~\bibnamefont{Kabalnov}},
  \bibinfo{author}{\bibfnamefont{K.}~\bibnamefont{Makarov}},
  \bibinfo{author}{\bibfnamefont{A.}~\bibnamefont{Pertsov}}, \bibnamefont{and}
  \bibinfo{author}{\bibfnamefont{E.}~\bibnamefont{Shckukin}},
  \bibinfo{journal}{J. Colloid Inteface Sci.} \textbf{\bibinfo{volume}{138}},
  \bibinfo{pages}{98} (\bibinfo{year}{1990}).

\bibitem[{\citenamefont{Egelhaaf et~al.}(1999)\citenamefont{Egelhaaf, Olsson,
  Schurtenberger, Morris, and {Wennerstr\"{\o}m}}}]{egelhaaf:ostrip}
\bibinfo{author}{\bibfnamefont{S.}~\bibnamefont{Egelhaaf}},
  \bibinfo{author}{\bibfnamefont{U.}~\bibnamefont{Olsson}},
  \bibinfo{author}{\bibfnamefont{P.}~\bibnamefont{Schurtenberger}},
  \bibinfo{author}{\bibfnamefont{J.}~\bibnamefont{Morris}}, \bibnamefont{and}
  \bibinfo{author}{\bibfnamefont{H.}~\bibnamefont{{Wennerstr\"{\o}m}}},
  \bibinfo{journal}{Phys Rev E} \textbf{\bibinfo{volume}{60}},
  \bibinfo{pages}{5681} (\bibinfo{year}{1999}).

\bibitem[{\citenamefont{Burnett et~al.}(1995)\citenamefont{Burnett, Chae, and
  Tam}}]{burnett:breaking}
\bibinfo{author}{\bibfnamefont{G.}~\bibnamefont{Burnett}},
  \bibinfo{author}{\bibfnamefont{J.}~\bibnamefont{Chae}}, \bibnamefont{and}
  \bibinfo{author}{\bibfnamefont{W.}~\bibnamefont{Tam}},
  \bibinfo{journal}{Phys. Rev. E} \textbf{\bibinfo{volume}{51}},
  \bibinfo{pages}{5788} (\bibinfo{year}{1995}).

\bibitem[{\citenamefont{Oswald}(1987)}]{oswald:bubble}
\bibinfo{author}{\bibfnamefont{P.}~\bibnamefont{Oswald}}, \bibinfo{journal}{J.
  Physique.} \textbf{\bibinfo{volume}{48}}, \bibinfo{pages}{897 }
  (\bibinfo{year}{1987}).

\bibitem[{\citenamefont{Zywocinski et~al.}(2000)\citenamefont{Zywocinski,
  Picano, and Geminard}}]{zywocinski:edge}
\bibinfo{author}{\bibfnamefont{A.}~\bibnamefont{Zywocinski}},
  \bibinfo{author}{\bibfnamefont{F.}~\bibnamefont{Picano}}, \bibnamefont{and}
  \bibinfo{author}{\bibfnamefont{J.~C.} \bibnamefont{Geminard}},
  \bibinfo{journal}{Phys. Rev. E} \textbf{\bibinfo{volume}{62}},
  \bibinfo{pages}{8133} (\bibinfo{year}{2000}).

\bibitem[{\citenamefont{Picano et~al.}(2000)\citenamefont{Picano, Holyst, and
  Oswald}}]{picano:meniscus}
\bibinfo{author}{\bibfnamefont{F.}~\bibnamefont{Picano}},
  \bibinfo{author}{\bibfnamefont{R.}~\bibnamefont{Holyst}}, \bibnamefont{and}
  \bibinfo{author}{\bibfnamefont{P.}~\bibnamefont{Oswald}},
  \bibinfo{journal}{Phys. Rev. E} \textbf{\bibinfo{volume}{62}},
  \bibinfo{pages}{3747 } (\bibinfo{year}{2000}).

\bibitem[{\citenamefont{Picano et~al.}(2001)\citenamefont{Picano, Oswald, and
  Kats}}]{picano:disjoining}
\bibinfo{author}{\bibfnamefont{F.}~\bibnamefont{Picano}},
  \bibinfo{author}{\bibfnamefont{P.}~\bibnamefont{Oswald}}, \bibnamefont{and}
  \bibinfo{author}{\bibfnamefont{E.}~\bibnamefont{Kats}},
  \bibinfo{journal}{Phys. Rev. E} \textbf{\bibinfo{volume}{63}},
  \bibinfo{pages}{21705} (\bibinfo{year}{2001}).

\bibitem[{\citenamefont{Cheng and Lemlich}(1983)}]{cheng:errors}
\bibinfo{author}{\bibfnamefont{H.~C.} \bibnamefont{Cheng}} \bibnamefont{and}
  \bibinfo{author}{\bibfnamefont{R.}~\bibnamefont{Lemlich}},
  \bibinfo{journal}{Ind. Eng. Fundam.} \textbf{\bibinfo{volume}{22}},
  \bibinfo{pages}{105 } (\bibinfo{year}{1983}).

\bibitem[{21}]{a:bubble}
\bibinfo{note}{This way of determining the real distribution of bubble size was
  determined by Lemlich \cite{cheng:errors}. The relation between the size
  distribution of bubbles in a 3D foam, $f(r)$, and with bubbles in a 2D
  slice,$F(r)$, is \[ F(r) = \left[ \int_{0}^{\infty} \frac{f(r)\textrm{d}r}{r}
  \right]^{-1} \frac{f(r)}{r} \] This can also apply to the distribution at the
  glass surface in the absence of counterbalancing local segregation.}

\end{thebibliography}
\end{document}